\documentstyle[preprint,prl,aps]{revtex}

\begin{document}
\title{Dynamical Aspects of Photoinduced Magnetism and Spin-Crossover phenomena in
Prussian Blue Analogs}
\author{Masamichi Nishino$^{1}$, Kamel Boukheddaden$^{2}$, Seiji Miyashita$^{3}$,
and Fran\c cois Varret$^{2}$}
\address{$^{1}${\it Computational Materials Science Center, National Institute}\\
for Materials Science, Tsukuba, Ibaraki 305-0047, Japan \\
$^{2}$ {\it Laboratoire de Magn\'{e}tisme et d'Optique, CNRS-Universit\'{e}}%
\\
de Versailles/St. Quentin en Yvelines 45 Avenue des Etats Unis, F78035\\
Versailles cedex, France. \\
$^{3}${\it Department of Physics, Graduate School of Science,}\\
The University of Tokyo, Bunkyo-Ku, Tokyo, Japan. }
\date{\today}
\maketitle
\begin{abstract}
We present dynamical properties of spin crossover compounds with photomagnetization, proposing a new model in which the spin-crossover phenomena and magnetic ordering are incorporated in a unified way. 
By using this model, the novel characteristics observed in Prussian blue analogs are qualitatively well reproduced. 
We investigate the time evolution of the magnetization and high 
spin fraction taking into account multi-time scales in a master equation approach for the magnetic relaxation, the lattice (electronic) relaxation, and the photoexcitation process. 
In particular, processes with different temperature cycles starting from the photoinduced saturated magnetic state are studied including 
the effect of photoirradiation.
In the temperature cycle in the low temperature region where the high spin state has a strong metastability, the magnetization almost exactly follows the quasi-static process.
On the other hand, when the temperature is raised above the region,
the dynamics of the spin state and the magnetization couple and exhibit various types of dynamical cooperative phenomena under time-dependent control of temperature and photoirradiation.
\end{abstract}


\section{Introduction}

Spin-crossover (SC) transitions between the low spin (LS) state and the high
spin (HS) state are induced by changes in temperature, pressure, photoirradiation, etc.\cite{Gutlich} because the LS or HS state is realized due to subtle balance of the interaction between the electrons and the ligand field.  
Phenomena originating from these transitions have attracted much attention because they would potentially produce various functional devices\cite{Kahn}. 
Among them, photoinduced phase transition is a current topic in the field of molecular metal complex solids. There, much effort has been dedicated to reversible controls of magnetic, electronic, structural properties, etc.\cite
{Gutlich,Kahn,Decurtins,Sato_1,Sato_2,Nasu,Desaix,Letard2,Hauser}. 
Molecular SC solids and Prussian Blue analogs (PBA) are typical materials that exhibit such properties.
In order to understand the phenomena, it is necessary to study the dynamical processes of the transitions of photoexcitation and the relaxation of electronic states, where various cooperative effects
play important roles. Thus, photoinduced non-equilibrium dynamics
is now an area of intense study \cite
{Koshihara,Goujon,Varret,Shimamoto,Sato3,Hashimoto,Photo2,Letard}.

It has been clarified that SC transitions are 
not assembly of independent transitions between two electronic states in individual molecules but cooperative phenomena in correlated molecules between a nonmagnetic (LS, diamagnetic) phase 
and a magnetic (HS, paramagnetic) phase \cite{Gutlich}.
Wajnflasz and Pick (WP) \cite{Wajnflasz1,Wajnflasz2} theoretically gave the ground of the mechanism of the cooperative transitions by using an Ising model with degeneracies. The WP-model and extended WP-models
have explained successfully static and dynamical properties of various cases of SC transitions\cite{Bousseksou1,Kamel1,Kamel2,Nishino1,Nishino2,Kawamoto}. 

The discovery of LIESST (lightinduced excited spin state trapping)
\cite{Gutlich,Decurtins} has led to a number of studies concerning relaxation during and after irradiation. 
The self-accelerated relaxation of HS fraction after irradiation and lightinduced thermal hysteresis, 
which are typical features in experiments, were reproduced by an Ising-like model
based on the WP-model with Arrhenius dynamics \cite{Kamel1}.
To reproduce the relaxation dynamics with a slowing down at long time, 
a treatment beyond mean-filed approaches is necessary\cite{Nishino2,Romstedt,Hoo}, 
where short range correlations (and sometimes long range correlations) between molecules and spatially non-uniform fluctuation are important. 

In a series of related experiments, a photoinduced magnetic phase transition
was observed in Co-Fe Prussian blue analogues (PBA)~\cite{Sato_1,Sato_2}.
At low temperatures, PBA shows the transition: Fe$^{{\rm II}}$(S=0) + Co$^{{\rm III}}$(S=0) 
$\rightarrow $ Fe$^{{\rm II}}$(S=1/2) + Co$^{{\rm III}}$(S=3/2) by photoirradiation.
In the equilibrium, the system is in the LS (the left hand side) state which is a nonmagnetic state.
By the photoirradiation, the system has a spontaneous magnetization where the spin state is in the HS
(the right hand side) state. Furthermore a reversible switching between the LS and HS states by irradiations with different frequencies was observed in this compound. 
There, the magnetic interaction between the high spins should play an important role in addition
to the lattice interaction which drives the LS-HS transition. Then, to study such multi-functional phenomena, it has become indispensable to establish a unified formulation which allows us to treat both SC transition and magnetic order-disorder transition simultaneously.

In order to investigate the reversible switching mechanism of magnetization, the Blume-Capel (BC)
model\cite{BC1,BC2}, which has a long-lived metastable ferromagnetic state 
in a parameter region, was studied\cite{Nishino3,Nishino4}. There, the dynamical process of the two-way switching was explained by focusing on the magnetic interaction 
and the multi-time scales of microscopic processes, e.g., the  LS-HS transition, spin flip, and photoexcitation.

For cooperated phenomena of the spin-crossover transition and magnetic ordering, 
we have proposed a unified model in which both the inter-molecular lattice interaction and the inter-molecular magnetic interaction are taken into account\cite{Kamel_static}. This generalized model includes both the WP and BC models as special cases and enables us to treat the spin-crossover phenomena and magnetic ordering in a unified way. 
We have studied static properties of the model, and obtained its phase diagram
as a function of the magnetic and lattice coupling parameters\cite{Kamel_static}.

In this paper, we study the dynamical properties of the generalized model.
In particular, we investigate the dynamical properties of the magnetization
and the HS fraction under temperature cycles after photoirradiation, and
we also study the process under photoirradiation, motivated by the available experiments on PBA\cite{Goujon,Varret,Shimamoto}.

In order to study the dynamical aspects, we consider a master equation with 
multi-time scales where the relative relation of the relaxation times of the processes, e.g., transition between the LS and HS states, spin flip, and photoexcitation, plays an important role. 
We 
adopt here Arrhenius dynamics to realize the slow dynamics of the LS$\leftrightarrow$HS transition
at low temperatures, while we use standard Glauber dynamics
for fast relaxation of the spin flip process. The process of photoexcitation is taken into account
by a local excitation process from the LS state (0) to the HS state with up or down magnetization ($\pm 1$) with the same probability. 
The strength of the irradiation is controlled in our formulation.  

In the present analysis, the following properties are clarified.
Because the interaction between spins is considered,   
the typical shapes of the increase of not only the HS fraction but also magnetization by photoirradiation 
at low temperatures are naturally observed as non-linear functions of irradiation time\cite{Goujon,Enachescu}.
Experimentally, it has been observed that the HS fraction and magnetization change drastically depending on the temperature cycling processes\cite{Goujon,Varret,Shimamoto}. 
Here, we theoretically study the mechanism of the dynamics under various temperature cycles. In a cycle, we first prepare saturated magnetization under photoirradiation, and after turning off the irradiation the system is warmed up to a certain temperature, and it is cooled down to the initial temperature. 
It is found that in the low temperature region where the excited 
high spin state has a strong metastability, the magnetization almost exactly follows the quasi-static process.
On the other hand, if the temperature is raised above the region,
the dynamics of the spin state and the magnetization couple and exhibit various types of dynamical cooperative phenomena under time-dependent control of temperature and photoirradiation.
Our model explains qualitatively the novel behavior found in the experiments\cite{Goujon,Varret,Shimamoto}, 
and also predicts new phenomena which give useful suggestions for related experiments of photoinduced hysteresis phenomena\cite{Desaix,Letard2}.

The layout of this paper is as follows.
In Sec. II, the model and method used in this study are explained. In Sec. III, dynamics in various temperature cycles are presented. The dynamical process under photoirradiation is given in Sec. IV.
Section V is devoted to summary and discussion.

\section{Model and Method}

We have to introduce a model with both the magnetic and elastic interactions for investigation of both SC transition and magnetic ordering.
The degeneracies of the HS and LS states play an important role in SC transitions \cite{Wajnflasz1,Wajnflasz2}. The effect of the degeneracies is expressed by a temperature dependent field\cite{Kamel1}. In the present work, we adopt the following model\cite{Kamel_static,eq}:
\begin{equation}
{\cal H}=-J\sum_{\langle i,j\rangle }S_{i}S_{j}-K\sum_{\langle i,j\rangle
}(S_{i}^{2}-\frac{1}{2})(S_{j}^{2}-\frac{1}{2})+(D-\frac{\ln g}{\beta }
)\sum_{i}(S_{i}^{2}-\frac{1}{2})-h\sum_{i} S_{i},\;\;\; S_{i}=0,1,-1,  \label{JKDeq}
\end{equation}
where $0$ denotes nonmagnetic state (LS state) and $1$ and $-1$ denote up spin and down spin states (HS state), respectively. Here, $J$ is the magnetic interaction, $D(>0)$ is the ligand-field energy, $g$ is the effective degeneracy ratio between the HS and LS states \cite{Kamel1}, $h$ is the magnetic field, and  $
\beta =\frac{1}{k_{{\rm B}}T}$. 
Although real PBA has a sub-lattice structure and a ferrimagnetic state, 
we simplified it to a simple cubic lattice and adopted ferromagnetic coupling 
($J>0$) with the nearest-neighbor $\langle i,j\rangle $ interaction between molecules (sites). The inter-molecular interaction originating from the elastic interaction is denoted by $K(>0)$.\ Here, we adopt an attractive interaction for the
same state (LS-LS or HS-HS). We set the parameters as $J=5$K, $K=300$K, $%
D=1200$K which give realistic values of transition temperatures for PBA compounds\cite{Goujon,Varret,Shimamoto}.
The effective degeneracy ratio $g$ is estimated to be 100$\sim$1000, which is attributed to the degree of freedom of spin, orbital, and lattice and molecular vibrations,
and is also temperature dependent \cite{Kamel1}. Here for simplicity, we consider $g$ as a constant and set $g=200$. 

To describe the dynamics of the model,
we adopt the stochastic dynamics described by the following master equation: 
\begin{equation}
\frac{d}{dt}P(S_{1}\cdots S_{k}\cdots S_{N},t)=\hat{L}(t)P(S_{1}\cdots
S_{k}\cdots S_{N},t),  \label{mastEqu}
\end{equation}
where $P$ is the probability of the state $\{S_{1}\cdots S_{k}\cdots S_{N}\}$
at time $t$, and $\hat{L}(t)$ denotes the time evolution operator which
consists of the transition probability per unit time $W_{k}(S_{k}\rightarrow
S_{k}^{\prime })$ at the $k$-th site, where $S_{k}$($S_{k}^{\prime }$) takes
a value $0$ , $1$, or $-1$.

In order to explain the dynamics of PBA, we have to introduce multi time scales for different relaxations\cite{Nishino4}. 
The time scale of spin inversion ($S=\pm 1$) is considered to be
much faster than that between LS and HS transition ($S=0$ and $|S|=1$)
because the latter is accompanied by a structural change. 
To describe the relaxation process between the HS and LS states adequately, 
we adopt Arrhenius dynamics\cite{Kamel1}. On the other hand, we adopt standard
Glauber type evolution between $S=1$ and $S=-1$ for fast relaxation of the spin flip process.
With these choices the difference between the dynamics of LS$\leftrightarrow$HS transition and spin flip can be effectively 
taken into account.

In Arrhenius dynamics, the transition probability is given in the form: 
\begin{equation}
W_{k}^{{\rm latt}}(S_{k}\rightarrow S_{k}^{\prime })=\frac{1}{2}\exp
(-\beta (E_{0}-E(S_{k}))),
\end{equation}
where $S_{k}$ is 0 ($\pm 1$) and $S_{k}^{\prime }
$ is $\pm 1$ (0), and $E(S_{k})$ is a local energy of eq. (\ref{JKDeq}). 
The quantity $E_{0}-E(S_{k})$ represents the effective energy barrier between
initial ($S_{k}$) and final ($S_{k}^{\prime }$) states and $E_{0}$ is set as 
$E_{0}=1000$K\cite{Kamel1,Nishino2}. 
We denote the time evolution operator $\hat{L}(t)$ of this process by $\hat{L}_{%
{\rm latt}}$. 

The transition probability of Glauber dynamics is given by 
\begin{equation}
W_{k}^{{\rm spin}}(S_{k}\rightarrow S_{k}^{\prime })=\frac{1}{2}%
(1-S_{k}\tanh \beta \Delta E),
\end{equation}
where $S_{k}=1$ ($-1$) and $S_{k}^{\prime
}=-1$(1) and $\Delta E=J\sum\limits_{i\neq k}S_{i}$. 
We denote this process
by $\hat{L}_{{\rm spin}}$. 

From the experimental data\cite{Goujon}, we assume that the photoirradiation 
causes a local excitation of LS $\rightarrow $ HS, i.e., 
$S_{k}=0\rightarrow S_{k}^{\prime }=\pm 1$ with the same probability:
\begin{equation}
W_{k}^{{\rm photo}}(0\rightarrow \pm 1)=\frac{1}{2}.
\end{equation}
We denote this process by $\hat{L}_{{\rm photo}}$.

The total time evolution including all processes is given by 
\begin{equation}
\hat{L}(t)=p_{{\rm latt}}\hat{L}_{{\rm latt}}(t)+p_{{\rm spin}}\hat{L}_{{\rm %
spin}}(t)+p_{{\rm photo}}\hat{L}_{{\rm photo}}(t), 
\end{equation}
where the coefficients are related to the relative strength of the processes.
Here, $p_{{\rm photo}}$ is related to the intensity of irradiation.
We control the intensity of irradiation by  $p_{{\rm photo}}$ 
under the condition: 
 $p_{{\rm latt}}+p_{{\rm spin}}+p_{{\rm photo}}=1$.
In this study, we take $p_{{\rm latt}}/p_{{\rm spin}}=1$. 
Other choices such as $p_{{\rm latt}}/p_{{\rm spin}}=0.5,2,$ etc. 
do not change the qualitative properties of the relaxation dynamics,
because the difference of the time scale is taken into account in the choices 
of $\hat{L}_{{\rm latt}}(t)$ and $\hat{L}_{{\rm spin}}(t)$.

The order parameters of the present model are the magnetization ($0 \leq m \leq 1$)
\begin{equation}
m=\frac{\langle (\sum_{i}S_{i})^{2}\rangle }{N^{2}}
\end{equation}
 and the high spin fraction ($0 \leq q \leq 1$)
\begin{equation}
q=\frac{\langle \sum_{i}S_{i}^{2}\rangle }{N},
\end{equation} 
where $N$ is the system size (set as $N=20^3$) and $\langle\cdots\rangle $ represents 
an average over the stochastic process.

\section{Dynamics in the temperature cycles.}

In this section, we study three kinds of temperature cycles (heating and cooling) after switching off irradiation to simulate the experiments\cite{Goujon,Varret}.
First, the system is irradiated ($p_{{\rm photo}}=0.01$) at a very low temperature ($T_{0}=5K$) under
a magnetic field ($h=0.1$) until the state becomes the saturated magnetic HS state ($q=1$ and $m=1$). 

\begin{figure}[tb]
\caption{ Time dependence of $m$ (solid line) and $q$ (dashed line) in the initial photoexcitation 
at $T_0=$5K. }
\label{initial_photo}
\end{figure}

In Fig.\ref{initial_photo} we show the time evolution of $m$ (solid line) and that of $q$ (dashed line) 
at $T_{0}$ under irradiation with $h=0.1$.  
The change of $m$ follows that of $q$ because at this low temperature 
the magnetic moments of excited HS states rapidly align in the direction of the magnetic field due to
the magnetic interaction.
Here, we find characteristic time evolution of $m$ and $q$. That is, at the beginning of the irradiation, 
$m$ and $q$ increase  slowly with concave shapes.  
Then, they are accelerated and finally saturate slowly with convex shapes. 
This time dependence is consistent with the experiments\cite{Goujon,Enachescu}. 
This result indicates the following. At the beginning of the photoirradiation, 
the irradiation makes local excited HS states but the cooperative effect tends to put them back to
LS states because almost all neighboring sites are in the LS state, 
and thus the growth of $q$ and $m$ is suppressed.
It should be noted that this initial suppression is due to the cooperative interaction
between LS sites.  In the present scheme, this effect is automatically taken into account.    
When the number of HS sites is increased, the cooperative effect between HS 
sites becomes significant, resulting in the acceleration of $q$ and $m$. 

After switching off the irradiation and field (setting $p_{{\rm photo}}=0$ and $h=0$), 
we change the temperature. 
The temperature is gradually increased up to $T_{{\rm max}}$, 
and then decreased to the initial temperature $T_{0}$ 
(i.e., $T_{0}\rightarrow T_{{\rm max}}\rightarrow T_{0}$). 
Here we change the temperature in 1K steps. 
At each temperature, first $1000$ Monte Carlo steps (MCS) are discarded for a transient time,
and then $1000$MCS is used to measure the quantities. 
Thus, the temperature dependence of the quantities denotes their time dependence. 
The following three cases are studied. 
(A) $T_{{\rm max}}=60$K, (B) $T_{{\rm max}}=250$K and (C) $T_{{\rm max}}=93$K.
We will see that, depending on the value of $T_{{\rm max}}$, the system 
shows qualitatively different dynamical behavior of $q$ and $m$.

\begin{figure}[tbp]
\caption{Time (temperature) dependence of $m$ (solid line) and $q$ (dashed line) in process A. 
The arrows with solid heads (bare heads) indicate the direction
of change of $m$ ($q$). }
\label{lowT}
\end{figure}

The time (and temperature) dependence of $m$ (solid line) and that of $q$ (dashed line) in process A are depicted in Fig. \ref{lowT}. 
The magnetization $m$ decreases with heating and disappears around $23$K. 
The magnetization in the cooling process almost exactly follows that in the warming process. 
This indicates that the dynamics of $m$ follows the quasi-static magnetization process. 
It is noted that $q(\simeq 1)$ is constant during the whole process. 
Although the free energy in the static mean-field theory for the equilibrium state as a function
of $m$ and $q$ does not have metastability for $q=1$, the strong metastability of $q=1$ appears 
as a result of Arrhenius kinetics with a relatively large $E_0$ as mentioned in the previous section.
Indeed, it is found that the metastability appears in the dynamical mean-field theory, where
the time evolution of the magnetization is given by the Van Hove equation: 
\begin{equation}
\dot{m}=-\frac{\partial f(m,q)}{\partial m}.
\end{equation}
As far as $q\simeq 1$ at low temperatures, the free energy is given by
\begin{equation}
f(m,1)\simeq p_{{\rm spin}} \left(\frac{1}{2}m^{2}-\frac{1}{\beta Jz} \ln (\cosh \beta Jzm) \right)
+{\rm const.},  \label{f_Van_Hove}
\end{equation}
where a double well potential of $m$ is realized.
Thus, under the metastability of $q$(=1), $m$ behaves quasi-statically and shows a typical temperature dependence of the second-order magnetic transition 
in both warming and cooling processes. 
Assuming $q=1$, the transition temperature is given by the Curie temperature of 3D Ising model on the simple cubic lattice\cite{Landau}: $T_{\text{C}}=4.5J=22.5$K.
This quasi-static behavior is consistent with experiments\cite{Goujon,Varret}. 

\begin{figure}[tbp]
\caption{Time (temperature) dependence of $m$ (solid line) and $q$ (dashed line) in process B.}
\label{highT}
\end{figure}

Next, we study a case with a high value of $T_{{\rm max}}$ ($T_{{\rm max}}=250$K, case B). 
Figure~\ref{highT} shows the time (and temperature) dependence of $m$ and $q$. 
Now we find self-accelerated (sigmoidal) relaxation of $q$ in the region 90-110K, above which $q$ disappears 
until the SC transition occurs at a higher temperature.
We find hysteresis behavior in the region 160-220K which is due to the SC transition.
The dependence of $q$ and that of $m$ on the temperature cycle have been found  
in the experiments of PBA\cite{Goujon,Varret,Shimamoto}.
In the process, 
the magnetic interaction has negligible contribution to the thermal transition of $q$ 
because of $J \ll K$. 
Thus the transition point of the spin crossover where $q=1/2$ is given by 
\begin{equation}
T_{{\rm cross}}=\frac{D}{k_{{\rm B}}\ln 2g}=200{\rm K}. 
\end{equation}
Because $T_{{\rm cross}}$ is less than the Ising phase transition point: 
$4.5\ >T_{\text{cross}}/\frac{K}{4}$, 
the system exhibits the first-order phase transition at $T_{{\rm cross}}$\cite
{Wajnflasz1,Wajnflasz2,Bousseksou1,Kamel1,Kamel2,Nishino1,Nishino2}. 
Thus, the center of the hysteresis depends on $D$ and $g$ and the width of the hysteresis depends on the ratio $K/D$ for fixed $g$ and the changing rate of the temperature, which indicates varieties of hysteresis loops in the experiments\cite{Shimamoto}. 
After arriving at $T_{{\rm max}},$ the system is cooled down. 
In the cooling process below the spontaneous thermal hysteresis region, $m$ and 
$q$ are zero, where the system is in the equilibrium state. 

\begin{figure}[tbp]
\caption{Relaxation of $q$ when the temperature raising is stopped at several temperatures (80, 85, 90, 95K) in process B.}
\label{sigmoidal_relax}
\end{figure}

If we stop warming at some temperature on the way to $T_{{\rm max}}$ from $T_{0}$ in process B and maintain the temperature, we observe characteristic sigmoidal relaxations of $q$ with a tail as depicted in Fig. \ref{sigmoidal_relax}.  
These relaxation curves with temperature dependence are also observed commonly 
in the related experiments\cite{Desaix,Hauser,Kamel1}.

\begin{figure}[tbp]
\caption{Time (temperature) dependence of $m$ (solid line) and $q$ (dashed line) in process C.}
\label{middleT}
\end{figure}

Let us consider the third case C ($T_{{\rm max}}=93 $K). 
Just before arriving at this $T_{{\rm max}}$ in the warming process, $q$
begins to relax. 
In Fig. \ref{middleT}, the time (and temperature) dependence of $m$ (solid line) and that of $q$ (dashed line) are plotted. The relaxation of $q$ is observable until $T\simeq 60$K in the cooling process. 
The quantitative relaxation profile depends on the changing rate of the temperature, 
but qualitative features are common.  
When the temperature is further decreased, $q$ freezes with a
non-saturated value of $q(<1)$. Because $q<1$, the substrate for the magnetic order is in a sense diluted, and the magnetic transition temperature in the cooling process is lower than the Curie temperature 
$T_{{\rm C}}$ for $q=1$. 
In this process, the fluctuations of $m$ and $q$ are large. 
Thus, we take average over ten independent simulations, 
and plot the average with error bars in Fig. \ref{middleT}.
This lowering of the magnetic transition temperatures was observed also in
the experiments\cite{Goujon,Varret}.

\section{Dynamics in the temperature cycle under photoirradiation}

\begin{figure}[tbp]
\caption{Effect of the irradiation ($p_{{\rm photo}}=0.0001$) on $m$ and $q$
in process B. (a) $q$ under continuous irradiation (dotted line) is compared with $q$ with no irradiation (dashed line). The arrows indicate the direction of change of $q$ in the former case. (b) $m$ (solid line) and $q$ (dotted line) in the low temperature region under continuous irradiation. }
\label{photo}
\end{figure}

In this section, we investigate the time (and temperature) dependence of the order parameters under photoirradiation, which would give useful suggestions to related experiments. 
We study the same temperature cycle as B under photoirradiation 
($p_{{\rm photo}}=0.0001$). 
In Fig.\ref{photo} (a), the time (and temperature) dependence of $q$ under the irradiation (dotted line) is shown in comparison with the data in the case of no irradiation (same dashed line as in Fig.3). 
Because photoexcitation enhances the HS state and suppresses the relaxation of $q$, the thermal hysteresis of $q$ in the region 160-220 K is shifted to the low temperature side. 
In addition, lightinduced hysteresis is observed at lower temperatures (10-120 K). The sigmoidal relaxation of $q$, which is the right branch of this hysteresis, starts at a higher temperature than in the case with no irradiation. This tendency is consistent
with the dynamical mean-field analysis of the Ising-like model\cite{Kamel1}.

In Fig. \ref{photo} (b), the time (and temperature) dependence of $m$ (solid line) and that of $q$ (dotted line) are shown, focusing on the low temperature region. 
In the warming process, $m$ disappears at a transition temperature which is the
same as $T_{{\rm C}}$. 
Indeed we find eq. (\ref{f_Van_Hove}) has no $p_{{\rm photo}}$-dependence 
when $q=1$ under irradiation ($p_{{\rm photo}} \neq 0$).
It is worth noting that in the cooling process, $m$ appears at a temperature lower than $T_{{\rm C}}$($=4.5J=22.5$K). 
The irradiation causes growth of the HS fraction $q$ around 30K and after a delay the magnetic order appears at $T_{\rm C}' \simeq 18$K. 
Before the full saturation of $q$, the HS sites have a diluted distribution ($q<1$). Thus, the critical temperature of the magnetization is lower than $T_{{\rm C}}$. 
At the critical point $T_{\rm C}'$, the high spin fraction is $q\simeq 0.6$ in Fig. \ref{photo} (b).

For this value of $q$, the magnetic order should appear at 0.53$T_{{\rm C}}$\cite{Landau2} in 
the randomly diluted lattice (percolated lattice). 
The present $T_{\rm C}'(\simeq 0.8T_{\rm C})$ is higher than this value, which should be attributed 
to correlated high spin states.
The nature of this correlation of the HS sites will be studied in detail elsewhere.

The present observation suggests that if the temperature at which 
the left branch of the photoinduced hysteresis of $q$ appears is near $T_{{\rm C}}$, we find the magnetic hysteresis driven by the photoinduced hysteresis of $q$ in the temperature cycle.  If $q$ appears at a much higher temperature, the temperature dependence of $m$ is the same as that in the warming process. 
We expect that this magnetic hysteresis will be found in experiments
by controlling the strength of irradiation.



\section{Summary and discussion}

We have presented a three-state model which allows us to describe both SC and magnetic properties of PBA with photoirradiation.
By using this model, the time (and temperature) dependence of magnetization ($m$) and that of the high spin fraction 
($q$) were observed under several temperature cycling processes.
In the case of the temperature cycle in the low temperature region 
($T_{{\rm max}}=60$K), the second order magnetic transition takes place at the Curie temperature ($T_{{\rm C}}\simeq 22.5$K) of the corresponding 3D Ising model in both the warming and cooling processes, because the strong metastable high spin state is realized 
($q\simeq 1$) in this temperature region. 
When $T_{{\rm max}}$ is high ($250$K), $q$ disappears showing the self-accelerated (sigmoidal) relaxation, and it appears again at a high temperature and
the thermal hysteresis of spin crossover is observed. In the cooling process, the system is in equilibrium (diamagnetic state) below the thermal hysteresis region. 
When $T_{{\rm max}}$ is moderate (93K), partial relaxation of $q$ is observed in the cooling process, and a magnetic phase transition occurs at a temperature lower than $T_{{\rm C}}$. 
There, the saturated $m$ is less than the fully saturated $m(=1)$ because the saturated 
$q$ is less than 1. Our results reproduce the experimental results\cite{Goujon,Varret} qualitatively, and thus we believe that the dynamics that we adopted here represents an important aspect of the real dynamics of PBA.

We have also investigated the dynamical properties in the temperature cycle under
irradiation. A shift of the sigmoidal relaxation of $q$ to a higher temperature and a shift of thermal hysteresis to a lower temperature are observed, 
because the photoexcitation makes high spin state more stable.
We found that the magnetic hysteresis caused by the photoinduced hysteresis of $q$ occurs under a special condition, which is expected to be observed in related experiments
in the future.

Very slow relaxation phenomena in photoinduced magnetization, such as
the different values of magnetization in ZFC and FC procedures like the spin-glass systems, 
have been reported in some PBA very recently\cite{Goujon,Varret}.
With the present model, those phenomena will be analyzed, which will be reported elsewhere. 

\acknowledgements
The present work was supported by Grant-in-Aid for Scientific Research from
MEXT of Japan, and Minist\`{e}re de l'Education Nationale and CNRS (PICS
France-Japan Program) of France.
The present work is also partially supported by
Grant-in-Aid from the Ministry of Education, Culture, Sports,
Science and Technology, and also by NAREGI Nanoscience Project, Ministry of
Education, Culture, Sports, Science and Technology, Japan.
The authors are grateful for financial support from the Grants. The numerical
calculation is supported by the supercomputer center of ISSP of Tokyo
university.


\begin{references}
\bibitem{eq}  
The Hamiltonian (\ref{JKDeq}) is equivalent to  $%
H=-J\sum_{i\neq j}s_{i}s_{j}-h\sum_{i}s_{i}-K\sum_{i\neq
j}s_{i}^{2}s_{j}^{2}+(D'-\frac{\ln g}{\beta })\sum_{i}s_{i}^{2}$ (BEG model when $g=1$) for static properties, used in a 
previous paper\cite{Kamel_static}. 
$g(=\frac{r}{u})$ reflects the $u$-fold LS state and the $r$-fold up spin (down spin) HS state. 
When $K=0$ and $g=1$, it corresponds to BC model.
Putting $\sigma _{i}=2Si^{2}-1$
with $J=0$, it reduces to WP model \cite{Bousseksou1,Kamel1}: ${\cal H}=-%
\frac{K}{4}\sum_{\langle i,j\rangle }\sigma _{i}\sigma _{j}+\frac{1}{2}%
\left( D-\frac{\ln 2g}{\beta }\right) \sum_{i}\sigma _{i}.\;\;\;$
\end{references}
\end{document}